\documentclass[twocols,grl]{agu2006}
\usepackage{graphicx}
\authorrunninghead{WESTBROOK ET AL.}

\titlerunninghead{ICE PARTICLE EVOLUTION}

\authoraddr{C. D. Westbrook,
Department of Meteorology, University of
Reading, Berkshire, RG6 6BB, UK.
(c.d.westbrook@reading.ac.uk)}

\begin{document}

%
%
%
 \setkeys{Gin}{draft=false}
%
%

%
%

\title{Theory and observations of ice particle evolution in cirrus using Doppler radar: evidence for aggregation}
%

%
%


\author{C. D. Westbrook, R. J. Hogan, A. J. Illingworth and E. J. O'Connor}
\affil{Department of Meteorology,
University of Reading, Berkshire, UK}

\begin{abstract}
Vertically pointing Doppler radar has been used to study the evolution of ice particles as they sediment through a cirrus cloud. The measured Doppler fall speeds, together with radar-derived estimates for the altitude of cloud top, are used to estimate a characteristic fall time $t_c$ for the `average' ice particle. The change in radar reflectivity $Z$ is studied as a function of $t_c$, and is found to increase exponentially with fall time. We use the idea of dynamically scaling particle size distributions to show that this behaviour implies exponential growth of the average particle size, and argue that this exponential growth is a signature of ice crystal aggregation.

\end{abstract}

%
%

%

\begin{article}

%
%

\section{Introduction}
The growth of ice crystals and aggregate snowflakes in clouds is a key process both for the development of precipitation (Jiusto and Weickmann 1973), and in terms of the effect such clouds have on climate (Houghton 2001). In this work, we use radar observations of deep cirrus to study the growth of ice particles as they sediment through the cloud.

Vertically-pointing measurements of radar reflectivity $Z$ and Doppler velocity $v_d$ were made using the 35~GHz (8.6~mm) `Copernicus' radar at the Chilbolton Observatory in southern England. At this wavelength, the overwhelming majority of cirrus-sized particles are within the Rayleigh regime where the backscattered intensity is proportional to the square of particle mass $m$:
\begin{equation}
\label{zeqn}
Z=\frac{36\left|K_{\mathrm{ice}}\right|^2}{0.93\pi^2\rho_{ice}^2}\times\int_0^{\infty} n(m)m^2\mathrm{d}m,
\end{equation}
where $\rho_{\mathrm{ice}}$ is the density of solid ice and $n(m)\mathrm{d}m$ is the number density of particles with mass between $m$ and $m+\mathrm{d}m$. The dielectric factor $K_{\mathrm{ice}}$ contains the information about the shape and dielectric strength of the particles: for spherical ice particles $K_{\mathrm{ice}}=\frac{\epsilon-1}{\epsilon+2}$ and the permittivity of ice $\epsilon$ at millimetre wavelengths is approximately 3.15 (Jiang and Wu 2004). The Rayleigh scattering approximation at 35~GHz is accurate to within 10\% for particles with a maximum dimension of 1~mm or less (Westbrook \textit{et al} 2006).

The Doppler velocity is $v_d=v_t+v_{\mathrm{air}}$, where $v_t$ is the $m^2$-weighted average terminal velocity of the particles and $v_{\mathrm{air}}$ is the vertical air motion.
We use these measurements to estimate a characteristic particle fall time $t_c$, which we define as the time for which the `average' particle (with terminal velocity $v_t$) has been falling. Note that the Doppler velocity is weighted by the reflectivity making it sensitive to the larger ice particles, and so our average fall time will also be weighted toward these large particles.
 Taking the cloud top height $h_{\mathrm{top}}$ to be the altitude at which there is no longer a detectable radar return, we calculate the fall time associated with height $h$ as:
\begin{equation}
\label{avtime}
t_c=\int_h^{h_{\mathrm{top}}}v_d^{-1}\mathrm{d}h.
\end{equation}
Given this new measure, we are in a position to investigate the evolution of the ice particles, by studying the variation of reflectivity $Z$ with increasing fall time $t_c$. The advantage of this method, as opposed to simply studying $Z$ as a function of height, is that $t_c$ represents the physical time for which the average ice particle has been falling to reach a given height $h$, allowing us to relate our results to theoretical models of ice particle growth. Note that we have implicitly assumed that the cloud is in a steady state, such that the properties of the ice particles at height $h$ do not change significantly over the length of time it takes a particle to fall from cloud top to cloud base (which is between 45 minutes and 2 hours for the cases shown here). Essentially this means that the cloud does not evolve significantly on this time scale and is advecting as a rigid body across the radar beam. We therefore apply our technique only to non-precipitating, well developed ice clouds where there is there is low wind shear.

\section{Cloud Data}
\begin{figure}
 \noindent\center{\includegraphics[width=8cm]{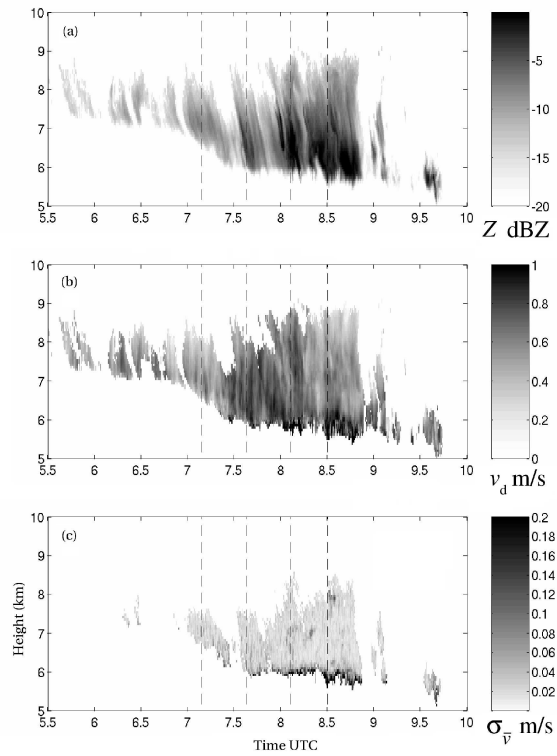}}
 \caption{\label{cloud} Radar time series of a cirrus cloud over Chilbolton on the $13^{\mathrm{th}}$ May 2004. Panels (a) and (b) show the reflectivity $Z$ and Doppler velocity $v_d$ respectively: both are averages over 30~s of data. Panel (c) shows the standard deviation $\sigma_{\overline{v}}$ of the 1-s average Doppler velocity for each 30-s period, indicating the variability in $v_d$.}
\end{figure}

Our case study is a cirrus cloud observed over Chilbolton on the $13^{\mathrm{th}}$ of May 2004. The temperature at cloud top (as forecast by the Met Office mesoscale model, Cullen 1993) was approximately $-40^{\circ}$C, and the cloud base was close to $-15^{\circ}$C; the average wind shear over the depth of the cloud was approximately $2 \mathrm{ms}^{-1}\mathrm{km}^{-1}$. Measurements of reflectivity and Doppler velocity were made and the time series of these observations is shown in figure \ref{cloud}. The radar gate length is 30~m (Illingworth \textit{et al} 2006). The values of $Z$ and $v_d$ are averages over periods of 30 seconds: in figure \ref{cloud}c we also show the standard deviation $\sigma_{\overline{v}}$ of the 1-s average Doppler velocity over each 30-s period, to indicate the small-scale variability in $v_d$. This measure allows the level of turbulence in the cloud to be assessed (Bouniol \textit{et al} 2003).

\begin{figure}
 \noindent\includegraphics[width=6.5in]{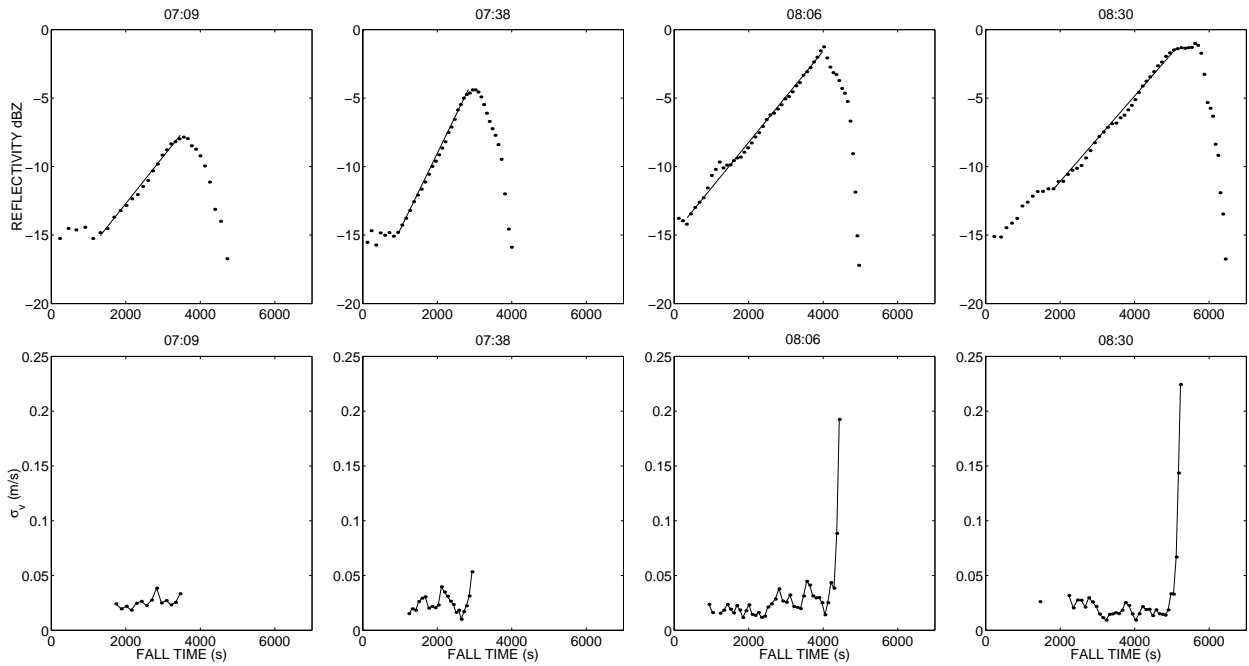}
 \caption{\label{zt} Four `snapshot' vertical profiles from the cirrus case, taken at 07:09, 07:38, 08:06, and 08:30 UTC. Each profile shown is the average of ten consecutive 30-s profiles. Top row is reflectivity in dBZ as a function of characteristic fall time (points). The solid line is intended to guide the eye, and indicates an exponential growth in $Z$ with $t_c$. Bottom row is $\sigma_{\overline{v}}$ as a function of $t_c$, which we use as an indicator of particle evaporation near cloud base.}
\end{figure}

Figure \ref{zt} shows four representative vertical profiles sampled from different portions of the cloud, indicated by the dashed lines on figure \ref{cloud}. Ten consecutive 30-s profiles were averaged over a period of $\sim$7 minutes in order to smooth out the variability caused by fall streaks in the data.
The highest detectable cloud pixel (corresponding to $\simeq-15$~dBZ) from the profile is taken as a measure of cloud top. The fall time at each height bin is calculated from the Doppler velocity profile as per equation \ref{avtime}, and we plot $Z$ as a function of $t_c$. From figure \ref{zt} we see that reflectivity increases rapidly with fall time (note the logarithmic dBZ units), which we interpret as rapid growth of the ice particles. This could potentially be occurring through a number of possible mechanisms: deposition of water vapour; aggregation via differential sedimentation of the ice particles; or collisions with supercooled drops (riming). In section 4 we show that it is likely that aggregation is the dominant growth mechanism. The increase in $Z$ appears to be exponential to a good approximation, and occurs for between 2500 and 5000 seconds in the profiles shown here. The slopes on the log scale vary between approximately $2.5\times10^{-3}$~$\mathrm{dBZ}$~$\mathrm{s}^{-1}$ and $5\times10^{-3}$~$\mathrm{dBZ}$~$\mathrm{s}^{-1}$, presumably depending on how much ice is being produced at cloud top. After this time there is a sharp turn over in the $Z(t_c)$ curves, and we attribute this to evaporation of the particles near cloud base. Such evaporation often results in increased air turbulence for which the particles themselves act as tracers, resulting in large variability in the Doppler velocity. In the earlier profiles (07:09 and 07:38 UTC) this was not evident; however, in the later profiles (08:06 and 08:30 UTC) the higher ice water content and time-integrated evaporative cooling triggered convective overturning and turbulence, and this is reflected in our observations (figures \ref{cloud} and \ref{zt}), which show a sudden increase in $\sigma_{\overline{v}}$ at approximately the same time as the turn over in $Z(t_c)$. 

\begin{figure}
 \noindent\center{\includegraphics[width=8cm]{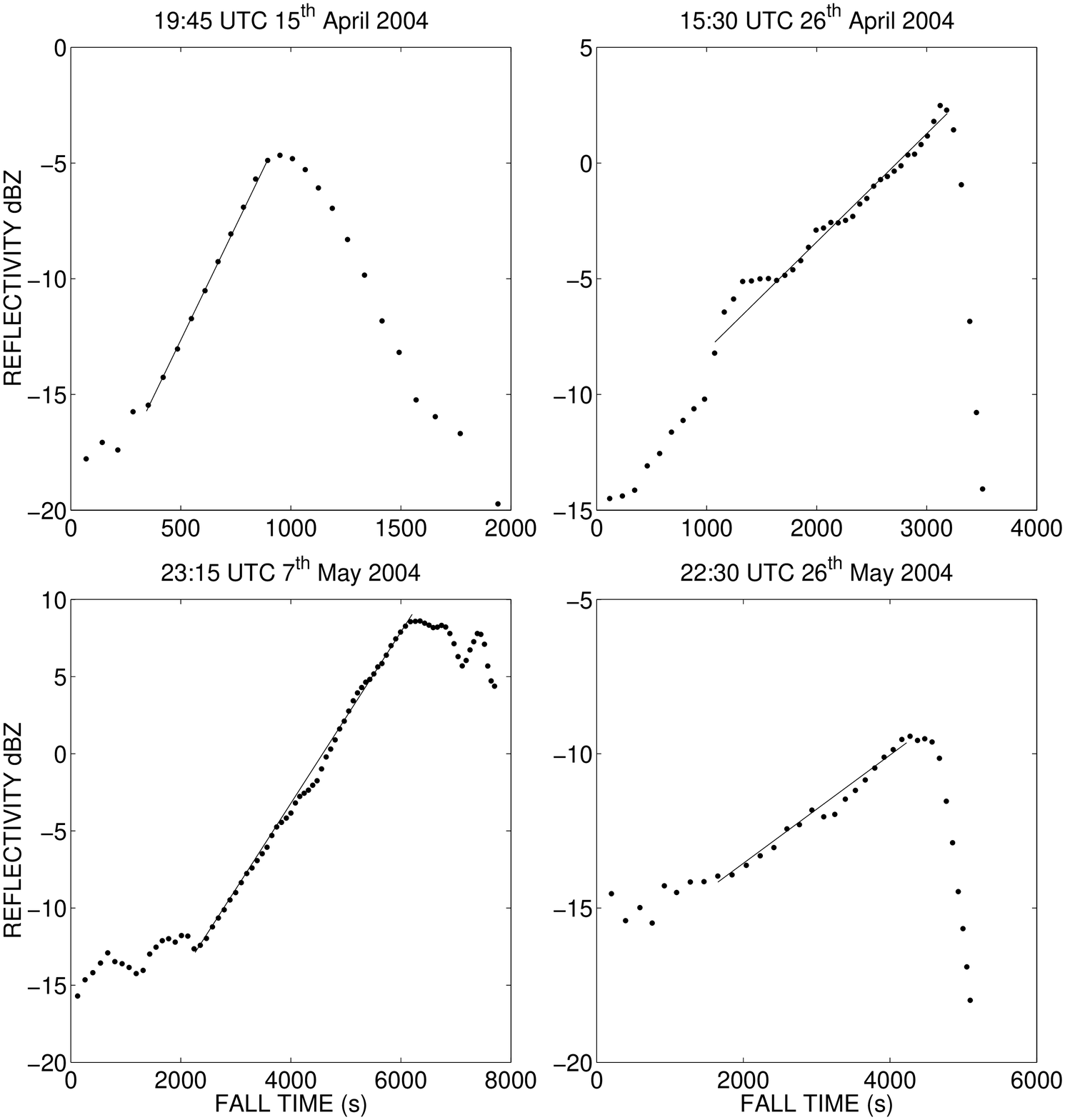}}
 \caption{\label{otherclouds}Vertical profiles of reflectivity from a sample of four more cirrus cases measured over Chilbolton during April and May 2004. Exact times and dates are indicated on the individual panels. All show an exponential growth of reflectivity with fall time over a significant portion of the cloud vertical profile.}
\end{figure}

Exponential growth has also been observed in a number of other cloud data sets, and four more example profiles from well developed non-precipitating ice clouds during April and May 2004 are shown in figure \ref{otherclouds}. This is an interesting feature of the data, and a robust one in the face of errors in $h_{\mathrm{top}}$: if $Z(t_c)$ is exponential, then even if we have underestimated the cloud top somewhat (on account of the limited sensitivity of the radar), this will merely correspond to an offset in the fall time, and the exponential shape of $Z(t_c)$ is still preserved. It is interesting to note that the transition from growth to evaporation is not always sharp as it is for the $13^{th}$ May profiles: we speculate that this may be the result of aggregation continuing to some extent within the evaporation layer.

\section{Scaling analysis}
Here we show how the reflectivity $Z$ is related to the average particle size. Scaling or `normalised' forms for the size distributions of both liquid and ice particles have been proposed in a number of recent articles (rain: Testud \textit{et al} 2001, Illingworth and Blackman 2002, Lee \textit{et al} 2005; ice: Field and Heymsfield 2003, Westbrook \textit{et al} 2004a,b, Delano\"{e} \textit{et al} 2005). The essence of these rescaling schemes is that the underlying shape of the distribution $\phi(m/\langle m\rangle)$ is the same throughout the vertical profile, but is rescaled as a function of the (increasing) average particle mass $\langle m\rangle$ as the particles grow:
\begin{equation}
\label{dseqn}
n(m)=\mathrm{IWC}\times\langle m\rangle^{-2}\phi\left(m/\langle m\rangle\right).
\end{equation}
where we have normalised by the ice water content IWC. The universal function $\phi$ is dimensionless. Equation \ref{dseqn} indicates that a single average particle mass $\langle m\rangle$ is sufficient to characterise the evolution of the particle size distribution (relative to the IWC or some other moment of the distribution), and this is key to our analysis.

An example of such a distribution is that assumed in the UK Met Office's Unified Model (Wilson and Ballard 1999). Mass $m$ and diameter $D$ are assumed to be in a power law relationship $m=a'D^b$, with an exponential distribution for particle diameter:
\begin{equation}
n'(D)=N_0\exp(-\Lambda D),
\end{equation}
where $n'(D)=n(m)\mathrm{d}m/\mathrm{d}D$. A single bulk prognostic variable is used for the ice particle mixing ratio and $N_0$ is parameterised to decrease with increasing temperature to mimic particle growth. The parameter $\Lambda$ is calculated from the predicted IWC and $N_0$, and is interpreted as a reciprocal average diameter $\langle  D\rangle^{-1}$ (eg. Brown \textit{et al} 1995). Within the framework (\ref{dseqn}) above, this distribution corresponds to:
\begin{equation}
\phi(x)=[b\Gamma(b+1)]^{-1}x^{(1-b)/b}\exp\left({-x^{1/b}}\right),
\end{equation}
where $x=m/\langle m\rangle$, and $\langle m\rangle=a'\langle D\rangle^b$.

Irrespective of what form is assumed for $\phi(x)$, a scaling relationship between different moments of the distribution may be found. The $k^{th}$ moment of the mass distribution is given by:
\begin{equation}
M_k=\int_0^{\infty} n(m)m^k\mathrm{d}m=\langle m\rangle^{k-1}\mathrm{IWC}\int_0^{\infty}\phi(x)x^k\mathrm{d}x.
\end{equation}
Note that $\int_0^{\infty}\phi(x)x^k\mathrm{d}x$ is a dimensionless constant.
Similarly, the radar reflectivity (\ref{zeqn}) is given by:
\begin{equation}
Z=\langle m\rangle\hspace{0.02in}\mathrm{IWC}\hspace{0.02in}\frac{36\left|K_{\mathrm{ice}}\right|^2}{0.93\pi^2\rho_{\mathrm{ice}}^2}\int_0^{\infty}\phi(x)x^2\mathrm{d}x.
\end{equation}
Combining these two equations we may relate $Z$ to an arbitrary moment $M_k$ of the distribution:
\begin{equation}
Z=\langle m\rangle^{2-k}\left(M_k\times\frac{36\left|K_{\mathrm{ice}}\right|^2}{0.93\pi^2\rho_{\mathrm{ice}}^2}\times\frac{\int_0^{\infty}\phi(x)x^2\mathrm{d}x}{\int_0^{\infty}\phi(x)x^k\mathrm{d}x}\right).
\label{scale}
\end{equation}
At this point we make a crucial assumption: that there is some moment of the distribution $k$ which is approximately constant through the vertical profile. In the case where aggregation is the dominant growth mechanism with a fixed production of ice mass at cloud top, one would expect the mass flux density of ice $\int_0^{\infty} n(m)mv(m)\mathrm{d}m$ to be constant. Mitchell (1996) indicated that a power law for ice particle fall speeds $v$ is a good approximation: $v(m)\propto m^c$, so for pure aggregation $k=1+c$. Similarly, where diffusional growth or riming is dominant, the total number flux of particles would be roughly constant and $k=c$ would be the conserved moment.
If this assumption holds then the bracketted expression $(\dots)$ in equation \ref{scale} is fixed through the vertical profile, and $Z\propto\langle m\rangle^{2-k}$.
Given our observations of exponential $Z(t_c)$ and the predicted power law between $Z$ and $\langle m\rangle$ above, we conclude that the average particle mass is growing exponentially with fall time.

\section{A Signature of Aggregation?}
We offer a possible explaination for the exponential growth of ice particles described above. Aircraft observations have indicated that aggregation is often the dominant growth mechanism for particles larger than a few hundred microns in cirrus clouds (Field and Heymsfield 2003), and it is these large particles which dominate the radar reflectivity.

Recently, Westbrook \textit{et al} (2004a,b) modelled ice particle aggregation by considering a rate of close approach between pairs of ice particles with masses $m$ and $m'$:
\begin{equation}
K=\frac{\pi}{4}\left(D_{max}+D_{max}'\right)^2\left|v-v'\right|,
\label{Kernel}
\end{equation}
where $v$ and $D_{max}$ are the associated fall speed and maximum dimension. Particles were picked according to the rate above, and traced along possible trajectories to accurately sample the collision geometries of the non-spherical ice particles. The fall speeds were prescribed in the vein of Mitchell (1996):
\begin{equation}
v\propto\frac{m^{\alpha}}{D_{max}},
\label{mitch}
\end{equation}
where the adjustable parameter $\alpha$ determines the drag regime (inertial flow $\alpha=\frac{1}{2}$; viscous flow $\alpha=1$).
One of the key results from these simulations was that the aggregates produced by the model had a power law relationship between mass and maximum dimension $m\propto D_{max}^b$, where the exponent is determined purely by the drag regime: $b=1/(1-\alpha)$ for $\alpha<\frac{2}{3}$. This relation is also backed up by a theoretical argument based on a feedback between the aggregate geometry and collision rate (Westbrook \textit{et al} 2004b). For large snowflakes, $\alpha\rightarrow\frac{1}{2}$ and $b\rightarrow2$, in good agreement with aircraft observations (eg. $b=1.9$, Brown and Francis 1995; $b=2.04$, Heymsfield \textit{et al} 2002).

In this study we are interested in the average ice particle growth rate, which is determined through the scaling of the collision kernel (\ref{Kernel}).
Given the above relationship between $a$ and $b$, and equations \ref{Kernel} and \ref{mitch}, we see that if one doubles the masses of the aggregating particles $m,m'$:
\begin{equation}
K(2m,2m')=2^{\lambda}K(m,m'),
\end{equation}
where $\lambda=\alpha+1/b=1$. This parameter controls the scaling of the particle growth rates and as such controls the growth of the average particle mass.

Van Dongen and Ernst (1985) have shown that the coagulation equation (Pruppacher and Klett 1997) has solutions with the same scaling form as (\ref{dseqn}), and predicts that the average particle mass grows according to the differential equation:
\begin{equation}
\frac{\mathrm{d}\langle m\rangle}{\mathrm{d}t_c}=w\langle m\rangle^{\lambda},
\label{dmdt}
\end{equation}
where $w$ is a constant. Given our prediction of $\lambda=1$ from the aggregation model:
\begin{equation}
\langle m\rangle\propto\exp(wt_c),
\label{mexpt}
\end{equation}
i.e. the prediction from aggregation theory is that average particle mass grows exponentially with fall time, in agreement with our observations. We note that the Van Dongen and Ernst analysis is for cases where total mass is conserved: however given the observed scaling behaviour (\ref{dseqn}) and a power law relationship between mass and fall speed, the case where mass flux density is conserved should yield the same result.

The growth of particles by diffusion of water vapour may also be described by a similar equation to (\ref{dmdt}). However in that case $\lambda=1/b$ and $w=4\pi C_0a^{-1/b}\times s/(A+B)$, where $C_0$ is the `capacitance' per unit diameter, $s$ is the supersaturation with respect to ice, and the terms $A$ and $B$ depend on temperature $T$ and pressure $P$ (Pruppacher and Klett 1997). For a given set of conditions $(s,T,P)$, the growth by deposition would be expected to increase slower with particle size than for aggregation, taking a power law form $\langle m\rangle\propto t_c^{b/(b-1)}$. In real clouds these conditions do not stay constant, and there is a correlation between increasing particle size and increased temperature and supersaturation, which could lead to a faster growth rate.
However, it would take a considerable conspiracy between these variables to obtain a constant exponential growth throughout such an extensive region of the cloud as is observed in our radar data. It also seems extremely unlikely that this correlation would be the same for all five cirrus cases shown in figures 2 and 3.
We note that there is a region of sub-exponential growth close to cloud top (small $t_c$) in some of the profiles in figure 3: we suggest that it is in this region, where the particles are small and falling slowly, that diffusional growth dominates.

It seems very unlikely that riming dominated the ice particle growth: a large number of supercooled drops throughout the depth of the cloud would be required for this to be the case. Given the cold temperatures in the cloud (between $-40^{\circ}$C and $-15^{\circ}$C as discussed in section 2), it is very unlikely that supercooled drops would persist on long enough time scales and in large enough quantities to dominate the growth over the 2.5~km or so for which we have observed $Z(t_c)$ to increase exponentially. We therefore discount deposition and riming, and assert that our observations are an indicator that aggregation is the dominant growth mechanism for the ice particles in these clouds.

\section{Discussion}
Doppler radar measurements of cirrus cloud were used to study the evolution of the ice particles sedimenting through it. The results indicate that in the cases studied the average ice particle mass grows exponentially with fall time, in agreement with the theoretical expectation for aggregation, and we believe that this is evidence that aggregation of ice crystals is the dominant growth mechanism for large particles in deep, well developed ice clouds.

Vertical profiles of reflectivity in ice have been much studied in order to estimate rainrates at the ground. Fabry and Zawadzki (1995) observed an approximately constant d(dBZ)/d$h$ gradient, and used this to rule out deposition as a growth mechanism. This may be linked to our cirrus observations; however their results were near the melting layer, and $Z$ was $\sim$20dB higher. We have compared profiles of dBZ-$h$ and dBZ-$t_c$ for our cirrus cases and find that while the dBZ-$t_c$ profiles are straight lines with a constant gradient, the dBZ-$h$ profiles have an appreciable curve to them. The fact that our analysis `straightens' these curved profiles is good evidence that our approach of using the Doppler velocity to estimate $t_c$ from $h$ is an appropriate one, and that aggregation is controlling the distribution of large ice particles.

The constant $w$ described in the aggregation theory above is directly related to the mass flux density, so measurements of the dBZ-$t_c$ slope may allow the derivation of this quantity, and the data could be combined with Doppler velocity measurements to estimate the ice water content. However, the sticking efficency of the ice particles (which we assume to be constant with particle size) is also a factor in $w$, and this is a parameter for which there are few reliable experimental estimates. For warmer, `stickier' ice crystals at temperatures above $-5^{\circ}$C this may be more feasible since the sticking efficiency should be close to unity.

We have assumed the ice particles fall vertically. In reality there is likely to be some horizontal shear, and this, combined with variability in ice production of the cloud-top generating cells results in visible fall streaks (see fig. 1). Size-sorting along the streaks (Bader \textit{et al} 1987) is a potential source of error in our analysis; however, by averaging the reflectivity profiles over $\sim$7 minutes of data we have been able to ameliorate it considerably.

Directions for future work are to make dual-wavelength radar measurements of cirrus in order to obtain a more direct estimate of particle size (Westbrook \textit{et al} 2006). This would help to pin down the dominant growth mechanism, allowing us to study moments other than $Z$, and analyse whether $k=1+c$ (aggregation) or $k=c$ (deposition, riming) is the moment conserved through the cloud. 
Aircraft observations (Field \textit{et al} 2005) have indicated a broadly exponential trend between $Z$ and temperature - it would be valuable to combine simultaneous radar and aircraft measurements to see if the exponential growth in $Z$ with $t_c$ is accompanied by exponential growth in $\langle m\rangle$ and increased concentrations of aggregates.
Also, further studies of other cirrus cases, both at Chilbolton and other radar sites, could be of interest to see how widespread the observed exponential trend is.

\begin{acknowledgments}
This work was funded by the Natural Environment Research Council (grant number NER/Z/2003/00643). We are grateful to the staff at the CCLRC Chilbolton Observatory, UFAM and the EU Cloudnet project (www.cloud-net.org), grant number EVK2-2000-00065. We are grateful to our two reviewers for their valuable comments and suggestions.
\end{acknowledgments}

%
%
%
%
%
%
%
%



\end{article}

%
%
%

\end{document}